\documentclass[runningheads]{llncs}

\usepackage[mobile]{eccv}

\usepackage{eccvabbrv}

\usepackage{graphicx}
\usepackage{booktabs}
\usepackage{multirow}
\usepackage{caption}
\usepackage[misc]{ifsym}

\usepackage[accsupp]{axessibility}  

\usepackage{hyperref}

\usepackage{orcidlink}

\definecolor{colorR1}{RGB}{237,109,82}
\definecolor{colorR2}{RGB}{96,176,119}
\definecolor{colorR3}{RGB}{66,145,247}

\begin{document}

\title{3DGS$^3$: Joint Super Sampling and Frame Interpolation for Real-Time Large-Scale 3DGS Rendering} 

\titlerunning{3DGS$^3$}




\makeatletter
\def\blfootnote{\xdef\@thefnmark{}\@footnotetext}
\makeatother

\author{%
  Yibo Zhao\inst{1*}, 
  Fan Gao\inst{1*}, 
  Youcheng Cai\inst{1(\textrm{\Letter})}, 
  Ligang Liu\inst{1}
}

\institute{
  University of Science and Technology of China
}
\authorrunning{Y.~Zhao et al.}

\maketitle
\blfootnote{\noindent$^{*}$Equal contribution.\\  \textsuperscript{\Letter}Corresponding author.}

\begin{abstract}
3D Gaussian Splatting (3DGS) enables high-quality real-time 3D rendering but faces challenges in efficiently scaling to ultra-dense scenes and high-resolution due to computational bottlenecks that limit its use in latency-sensitive applications. Instead of optimizing the splatting pipeline itself, we propose \textbf{3DGS$^3$}, a unified post-rendering framework that jointly performs super sampling and frame interpolation through differentiable processing of low-resolution outputs to achieve both high-resolution and high-frame-rate rendering. Our \textbf{Gradient\- \-Aware Super Sampling (GASS)} module leverages the continuous differentiability of 3DGS to extract image gradients that guide a GRU-based refinement network to enable high-fidelity super sampling. Furthermore, a \textbf{Lightweight Temporal Frame Interpolation (LTFI)} module based on a compact U-Net-like backbone fuses temporal and differentiable spatial cues from consecutive frames to synthesize temporally coherent intermediate frames. Experiments on public datasets demonstrate that 3DGS$^3$ achieves superior rendering efficiency and visual quality when compared with state-of-the-art methods and remains compatible with existing 3DGS acceleration techniques. The code will be publicly released upon acceptance.
\end{abstract}

\section{Introduction}
\label{sec:intro}

3D Gaussian Splatting (3DGS) \cite{kerbl20233d} has recently emerged as a powerful representation for real-time, high-quality 3D reconstruction by modeling scenes as a collection of Gaussian primitives. Although 3DGS enables real-time rendering across various domains, its performance is constrained by the high parameter count and inherent algorithmic inefficiencies, thereby limiting its broader adoption in latency-sensitive applications such as virtual reality (VR) \cite{franke2025vr, jiang2024vr}, augmented reality (AR) \cite{li2023instant}, and mobile devices \cite{lee2024gscore}.

Recent efforts to improve 3D Gaussian Splatting (3DGS) rendering efficiency can be broadly categorized into three major directions: (1) compact representations: methods such as GES~\cite{hamdi2024ges} and DisC-GS~\cite{qu2024disc} modify Gaussian primitives, through generalized exponential splats or boundary-aware formulations, to achieve more compact representations. (2) Gaussian reduction: methods such as LightGaussian~\cite{fan2024lightgaussian} and Mini-Splatting~\cite{fang2024mini} prune redundant splats based on spatial distributions to maintain visual quality with fewer primitives. (3) rendering acceleration: AdR-Gaussian~\cite{xzwang2024adrgaussian} introduces adaptive-radius filtering, while GSCore~\cite{lee2024gscore} leverages hardware-level optimizations to accelerate rasterization. Despite these advances, a fundamental bottleneck persists: achieving real-time rendering remains challenging, particularly when scaling to ultra-dense scenes or ultra-high-resolution outputs (e.g., millions of splats or 4K rendering). These factors impose substantial computational overhead, preventing 3DGS from achieving the responsiveness required for interactive experiences.

\begin{figure}[t] 
\centering 
\includegraphics[width=0.95\textwidth]{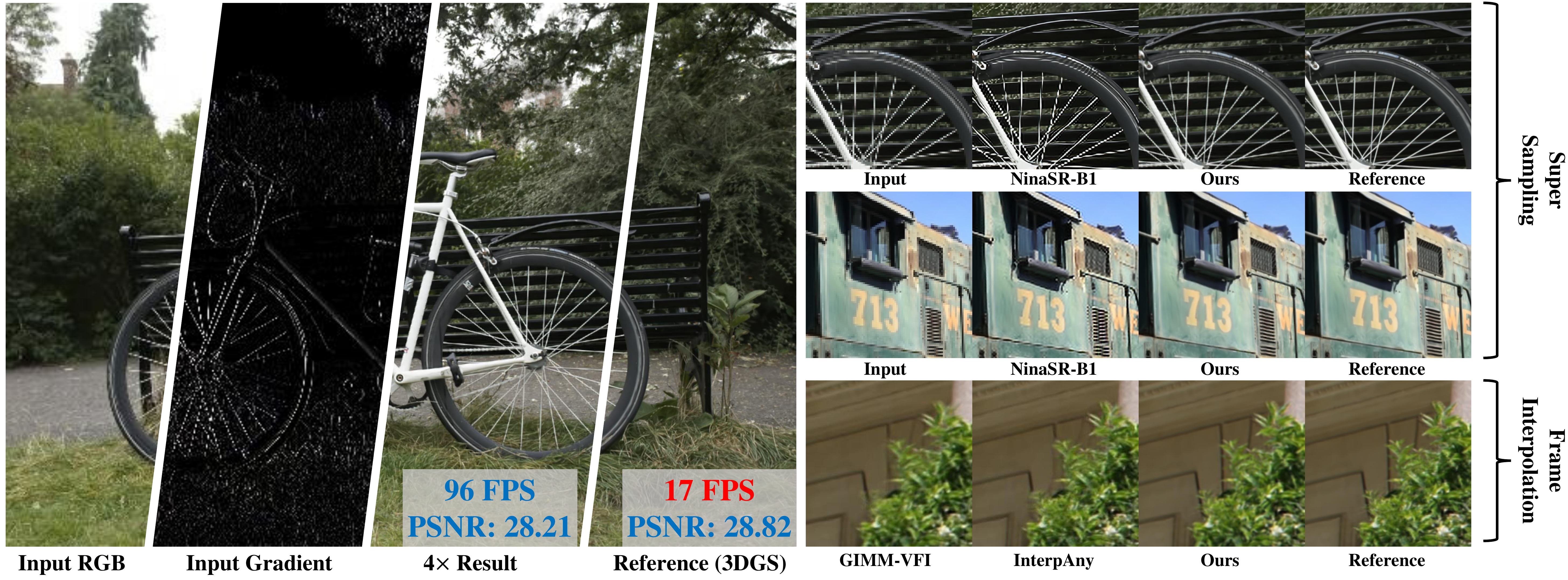}
\caption{We propose a novel post-rendering framework to accelerate 3DGS rendering. Leveraging spatio-temporal information, it jointly performs super sampling and frame interpolation over rendered outputs to achieve efficient acceleration while preserving high rendering quality.}
\label{fig:teaser-half}
\end{figure}

Rather than further optimizing the splatting pipeline itself, we investigate ways to accelerate 3DGS from the perspective of post-rendering, drawing inspiration from modern real-time graphics pipelines. In conventional rendering, commercial systems such as DLSS \cite{liu2020dlss}, XeSS \cite{intel2022xess}, and FSR \cite{amd2021fsr} achieve high-quality real-time performance by employing super sampling and frame interpolation techniques to reconstruct 4K resolution or high frame rates from low-resolution inputs. This paradigm shift motivates us to explore whether a similar approach can be effectively applied to 3DGS. However, directly integrating such methods into 3DGS is challenging, as a naïve integration often degrades quality and cannot accommodate the inherently continuous nature of Gaussian representations.

To address these challenges, we propose 3DGS$^3$, a unified framework that jointly performs \textbf{S}uper \textbf{S}ampling and frame interpolation for real-time large-scale 3DG\textbf{S} rendering. Inspired by super sampling and frame interpolation techniques in conventional rendering pipelines, 3DGS$^3$ performs differentiable post-rendering on low-resolution 3DGS outputs, thereby achieving both high-resolution and high-frame-rate rendering without modifying the underlying splatting pipeline (see Fig. \ref{fig:teaser-half}). Specifically, we design a \textbf{Gradient-Aware Super \ Sampling \ (GASS)} that leverages the continuous differentiability of Gaussian splats to extract image gradients serving as auxiliary cues. These gradient features guide image interpolation and a GRU-based refinement network to efficiently reconstruct high-resolution renderings while preserving perceptual fidelity. In addition, a \textbf{ Lightweight Temporal Frame Interpolation (LTFI)} based on a compact U-Net-like backbone fuses temporal and differentiable spatial cues from consecutive frames to synthesize intermediate images, thereby substantially enhancing rendering frame rates. Extensive experiments on real-world datasets demonstrate that 3DGS$^3$ achieves superior rendering efficiency and visual fidelity when compared with existing 3DGS acceleration and upscaling approaches.

Our main contributions are summarized as follows:
\begin{itemize}
\item \textbf{A new paradigm for 3DGS acceleration.} To the best of our knowledge, this work presents the first post-rendering framework that jointly performs super sampling and frame interpolation for 3DGS rendering.

\item \textbf{Gradient-aware super sampling.} We exploit the differentiable properties of Gaussians to extract continuous image gradients, which guide image interpolation and refinement to achieve efficient and high-quality rendering.

\item \textbf{Lightweight temporal frame interpolation.} We propose a lightweight U-Net-like interpolation module that fuses temporal and differentiable spatial cues to generate high-quality intermediate frames.
\end{itemize}

\section{Related Work}
\textbf{Accelerating 3DGS Rendering.}
3D Gaussian Splatting (3DGS) \cite{kerbl20233d} offers a differentiable and explicit representation that enables real-time, high-quality rendering and has been applied to autonomous driving, human avatars, and dynamic scenes. In this section, we review efforts to accelerate 3DGS rendering, which primarily focus on three aspects: (1) compact representations, (2) Gaussian reduction, and (3) rendering acceleration.

\textit{Compact representations.} These methods design more expressive and compact primitives to reduce redundancy. GES \cite{hamdi2024ges} introduces generalized exponential splatting, while DisC-GS \cite{qu2024disc} models depth and color discontinuities using a discontinuity-aware Gaussian for more realistic rendering. BG-Triangle \cite{wu2025bg} vectorizes Gaussians with Bézier triangles, and DRK \cite{huang2025deformable} adopts deformable radial kernels to improve modeling flexibility. Scaffold-GS \cite{lu2024scaffold} uses anchor-based adaptive parameters, and LocoGS \cite{shin2025locality} applies locality-aware compression for faster rendering.

\textit{Gaussian reduction.} These methods prune redundant primitives to reduce computational cost. EAGLES \cite{girish2024eagles} combines quantized embeddings and pruning for a ten- to twenty-fold reduction. LightGaussian \cite{fan2024lightgaussian} compresses unbounded Gaussians for over 200 fps rendering, while Mini-Splatting \cite{fang2024mini} and PUP-3DGS \cite{hanson2025pup} perform budgeted and uncertainty-based pruning. RadSplat \cite{niemeyer2025radsplat} uses radiance-aware selection, achieving 900 fps rendering.

\textit{Rendering acceleration.} Other methods enhance pipeline efficiency through hardware and rasterization optimizations. ADR-Gaussian \cite{xzwang2024adrgaussian} adapts Gaussian radius to reduce overdraw, Speedy-Splat \cite{hanson2025speedy} exploits sparsity of pixels and primitives for fast splatting, FlashGS \cite{feng2025flashgs} optimizes GPU pipelines for large-scale scenes, and GSCore \cite{lee2024gscore} introduces hardware-level optimizations.

Despite these advances, real-time rendering under ultra-dense or ultra-high-resolution settings remains challenging. UpscaleGS \cite{niedermayr2025upscaling3dgs} integrates super sampling into 3DGS but requires scene-specific retraining. In contrast, our modular, scene-agnostic post-rendering framework performs differentiable processing on low-resolution 3DGS outputs to generate high-resolution and high-frame-rate results, remaining complementary to existing accelerations.

\textbf{Real-time Super Sampling.}
Traditional anti-aliasing techniques such as MSAA \cite{akeley1993reality}, MLAA \cite{reshetov2009morphological}, and SMAA \cite{jimenez2012smaa} enhance image quality through spatial averaging or edge-aware filtering, while temporal super sampling \cite{yang2009amortized} reuses samples across frames. However, these heuristic methods have difficulty preserving fine details. Neural approaches like DLSS \cite{liu2020dlss} and XeSS \cite{intel2022xess} leverage temporal and spatial cues for reconstruction, along with NSRR \cite{xiao2020neural}, NSRD \cite{li2024neural}, and FuseSR \cite{zhong2023fusesr}, which enhance temporal stability through end-to-end learning. Recently, RDG \cite{zheng2025efficient} adopts decoupled G-buffer guidance to improve temporal coherence and overall rendering consistency.

\textbf{Real-time Frame Generation.}
Early frame synthesis relied on temporal reprojection and 3D warping \cite{mark1997post, scherzer2012temporal, amd2022fsr}, reusing past frames to increase frame rates but suffering from ghosting and occlusion issues. Neural methods \cite{guo2021extranet, li2022future, briedis2021neural, wu2023adaptive, briedis2023kernel, wu2023extrass} instead learn motion-aware features to predict or interpolate frames, achieving higher temporal consistency, although they depend on rasterized G-buffers, limiting applicability to 3DGS. Complementary progress in video frame interpolation, ranging from phase- or kernel-based \cite{meyer2015phase, niklaus2017video, ding2021cdfi, lee2020adacof} to optical flow–based approaches \cite{liu2017video, bao2019depth, kalluri2023flavr, guo2024generalizable}, further motivates our differentiable post-rendering design for temporal upsampling.

\section{Preliminaries}
\textbf{3D Gaussian Splatting.} 3DGS \cite{kerbl20233d} represents a 3D scene as a comprehensive set of anisotropic Gaussian primitives. Each Gaussian primitive is defined as:
\begin{equation}
G(\mathbf{x}) = \exp\left[-\frac{1}{2}(\mathbf{x} - \mathbf{x}_c)^T \mathbf{\Sigma}^{-1}(\mathbf{x} - \mathbf{x}_c)\right],
\end{equation}
where $\mathbf{x}_c$ represents the Gaussian center and $\mathbf{\Sigma}$ is the covariance matrix, which can be factorized as $\mathbf{\Sigma} = \mathbf{R}\mathbf{S}\mathbf{S}^{T}\mathbf{R}^{T}$for a rotation matrix $\mathbf{R}$ and a scaling matrix $\mathbf{S}$. Subsequently, the Gaussians are projected into the image space via EWA Splatting \cite{zwicker2002ewa}, which enables efficient rasterization-based rendering. The resulting 2D Gaussians are depth-sorted, and pixel colors are computed using $\alpha$-blending:
\begin{equation}
        I = \sum_{i \in N} c_i \alpha_i T_i =  \sum_{i \in N} c_i \alpha_i\prod_{j=1}^{i-1}(1 - \alpha_j),
\end{equation}
where $\alpha_i$ is computed by the Gaussian multiplied with the opacity and $c_i$ is the view-dependent color predicted from spherical-harmonics (SH) coefficients.

\textbf{3DGS Image Gradients.} UpscaleGS \cite{niedermayr2025upscaling3dgs} proposes utilizing the analytical image-space gradients of Gaussians, obtained from the exact derivatives of the smooth, continuous signal. Assuming $N$ Gaussians represent the scene, the rendered image $I(x,y)$ is differentiable with respect to the 2D spatial coordinates $(x, y)$:
\begin{equation}
\begin{aligned}
& \frac{\partial I(x, y)}{\partial x}=\sum_{i=1}^N c_i(\frac{\partial T_i}{\partial x} \alpha_i+T_i \frac{\partial \alpha_i}{\partial x}), \\
& \frac{\partial I(x, y)}{\partial y}=\sum_{i=1}^N c_i(\frac{\partial T_i}{\partial y} \alpha_i+T_i \frac{\partial \alpha_i}{\partial y}).
\end{aligned}
\end{equation}
Here, the differential of $T_i$ can be recursively expanded as:
\begin{equation}
     \frac{\partial T_i}{\partial x} = - \frac{\partial \alpha_{i-1}}{\partial x}T_{i-1} + (1 - \alpha_{i-1}) \frac{\partial T_{i-1}}{\partial x},
\end{equation}
and the computation of $\frac{\partial T_i}{\partial y}$ follows an analogous form. The image-space gradients are obtained by blending the weighted per-Gaussian gradients.

\section{Method}
\begin{figure*}[t] 
\centering 
\includegraphics[width=1\textwidth]{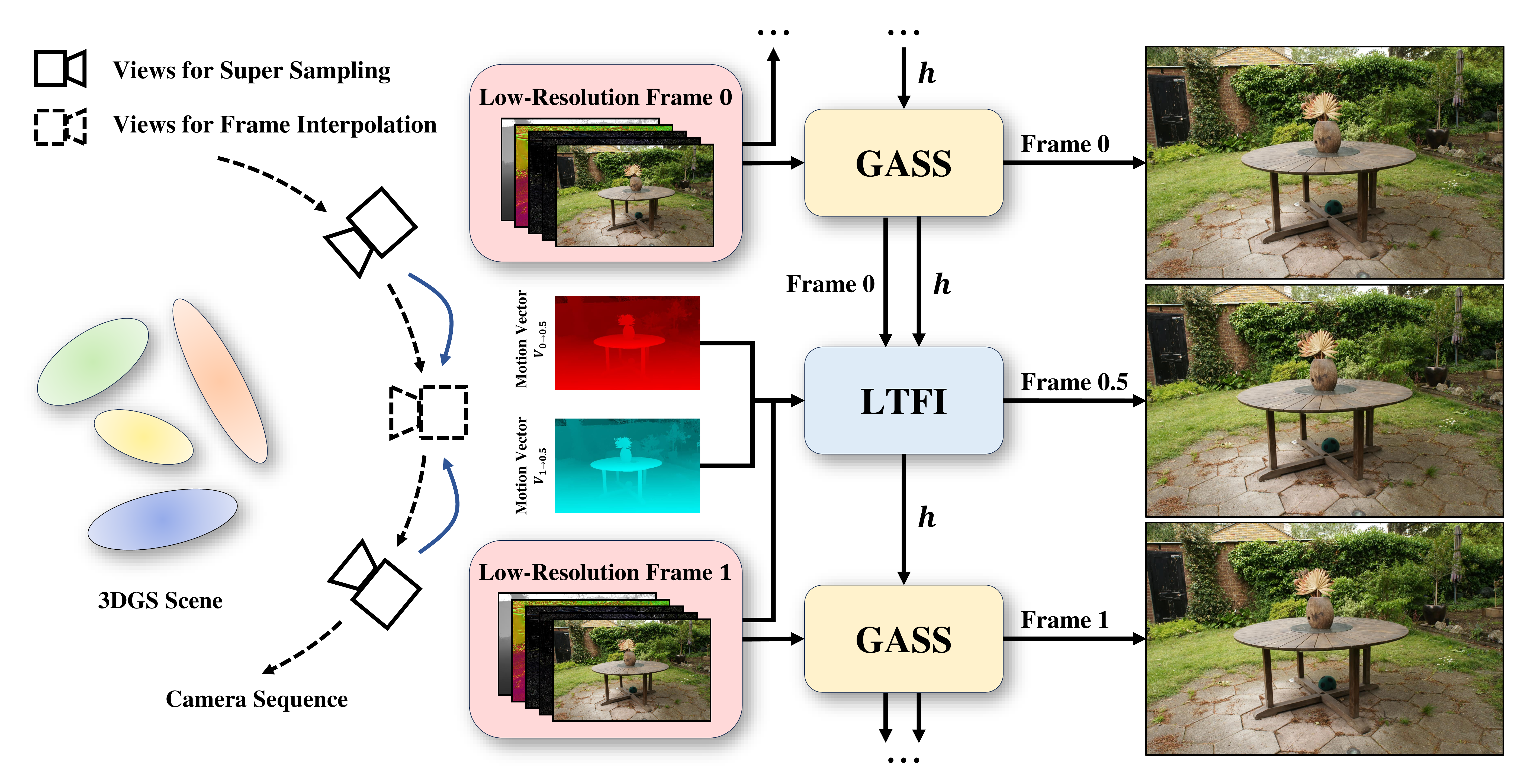}
\caption{Overview of our approach. (1) The \textbf{Gradient-Aware Super Sampling (GASS)} module processes the low-resolution frames rendered by 3DGS by combining them with image gradients and the historical feature to generate high-resolution images. (2) The \textbf{Lightweight Temporal Frame Interpolation (LTFI)} module uses the high-resolution image together with the subsequent low-resolution frame to interpolate intermediate frames.}
\label{fig:pipeline-overview}
\end{figure*}

In this section, we introduce \textbf{3DGS$^3$}, a unified framework that jointly performs super sampling and frame interpolation to accelerate 3DGS rendering. The overall structure of our pipeline is illustrated in Figure \ref{fig:pipeline-overview}. First, we present a \textbf{Gradient-Aware Super Sampling (GASS)} module that leverages the differentiable nature of 3DGS-rendered images to achieve high-quality super sampling. Furthermore, we propose a \textbf{Lightweight Temporal Frame Interpolation (LTFI)} module based on a compact U-Net-like backbone to reduce computational cost by effectively leveraging temporal and differentiable spatial cues. The two modules work collaboratively to reduce the average rendering time per frame, thereby improving overall rendering efficiency.

\begin{figure}[t] 
\centering 
\includegraphics[width=1\textwidth]{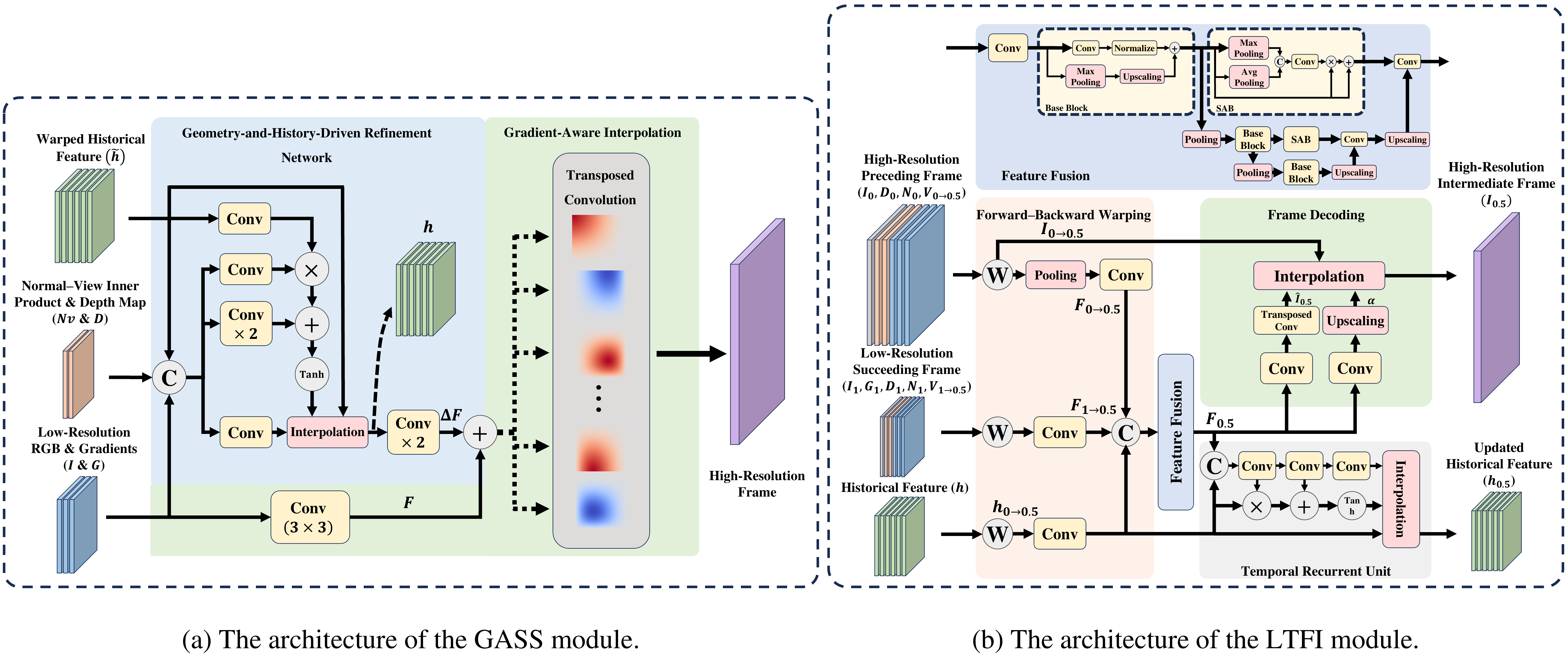}
\caption{The detailed architecture of the (a) GASS module, which consists of a gradient-aware interpolation module and a geometry-temporal recurrent refinement network; and (b) LTFI module, which consists of a Forward–Backward Warping block, a Feature Fusion Network, and a Temporal Recurrent Unit.}
\label{fig:sr}
\end{figure}

\subsection{Gradient-Aware Super Sampling}
Unlike UpscaleGS \cite{niedermayr2025upscaling3dgs}, which employs image-space gradients to determine a bicubic spline function fitted to neighboring pixels around the target pixel for interpolation, we propose a GASS network that that more effectively and robustly leverages geometric and historical priors to achieve superior super sampling performance, consisting of two main components: (1) {Gradient-aware interpolation} and (2) {Geometry-temporal recurrent refinement}.

\textbf{Gradient-Aware Interpolation.} Inspired by Hermite interpolation \cite{kincaid2009numerical}, we propose a novel gradient-aware interpolation method. Hermite interpolation incorporates both function values and their corresponding derivatives at the interpolation nodes. For each interpolation node, the polynomial is constrained to match the function values and its derivatives up to a specified order of the original function.

For a given pixel $(x_0, y_0)$, we consider it along with its eight adjacent pixels, forming a $3\times3$ block of interpolation nodes. Although two-dimensional interpolation schemes may vary depending on node distribution and derivative availability, we consider a bivariate polynomial $f(x_0,y_0)$ that satisfies the following conditions:
\begin{equation}
\begin{aligned}
f(x_0+i,y_0+j) &= I(x_0+i,y_0+j),\\
\frac{\partial}{\partial x}f(x_0+i,y_0+j) &= \frac{\partial}{\partial x}I(x_0+i,y_0+j),\\
\frac{\partial}{\partial y}f(x_0+i,y_0+j) &= \frac{\partial}{\partial y}I(x_0+i,y_0+j),
\end{aligned}
\end{equation}
where $i,j \in \{-1,0,1\}$. Thus, the interpolated image $\hat{I}$ can be viewed as evaluating different local polynomials $f(x,y)$ at their corresponding spatial positions. For simplicity, we define that under an $S\times$ super sampling setting, the $3\times3$ block centered at $(x_0,y_0)$ in $I$ corresponds to an $S\times S$ block centered at $(\hat{x}_0,\hat{y}_0)$ in $\hat{I}$:
\begin{equation}
I_{3\times3}(x_0, y_0)\leftrightarrow \hat{I}_{3\times3}(\hat{x}_0, \hat{y}_0). 
\end{equation}
Consequently, each pixel within the block can be expressed as a linear combination of the low-resolution intensities and their spatial gradients, weighted by a coefficient matrix $\mathbf{M}$:
\begin{equation}
\mathrm{Vec}(\hat{I}_{S\times S}(\hat{x}_0, \hat{y}_0)) = \mathbf{M} \begin{bmatrix}
    \mathrm{Vec}(I_{3\times3}(x_0, y_0)) \\
    \mathrm{Vec}(\frac{\partial}{\partial x}I_{3\times3}(x_0, y_0))\\
    \mathrm{Vec}(\frac{\partial}{\partial y}I_{3\times3}(x_0, y_0))
\end{bmatrix}
\label{eq:M}
\end{equation}
where $\mathrm{Vec}(\cdot)$ denotes the column-wise vectorization operator. This operation can be implemented by cascading a convolution layer with a transposed-convolution layer:
\begin{equation}
\hat{I} = \operatorname{TConv}(F), F = \operatorname{Conv}(I),
\label{eq:ctc}
\end{equation}
where $\operatorname{Conv}(\cdot)$ denotes a $3\times3$ convolution layer and $\operatorname{TConv}(\cdot)$ denotes an $S\times S$ transposed-convolution layer. We interpret $F$ as the coefficient vector in the polynomial space, while $\operatorname{TConv}(\cdot)$ serves as a linear functional from the dual space, mapping elements of the polynomial space to pixel-wise intensities. Notably, the number of channels in $F$ is chosen to exceed its theoretical minimum dimension $\min\{S^2, 27\}$, ensuring that $\mathbf{M}$ in Eq.~\eqref{eq:M} possesses at least full column or row rank. This design choice enables the network to capture higher-order relationships.

\textbf{Geometry-Temporal Recurrent Refinement.}
\label{sec:irn}
While the introduction of gradients enhances the recovery of fine image details, these gradients are themselves derived from the partial derivatives of the original continuous image function. Consequently, they remain constrained by the sampling frequency, making it difficult to avoid aliasing artifacts. Recent works~\cite{zhong2023fusesr, wu2023adaptive} have demonstrated that incorporating G-buffers and temporal cues can significantly improve reconstruction quality. Motivated by this, we propose a Geometry-Temporal Recurrent Refinement Network to further enhance the interpolated outputs.

As illustrated in Fig.~\ref{fig:sr} (a), the proposed network employs a modified gated recurrent unit (GRU)~\cite{cho2014learning} architecture. The network takes as input the current low-resolution frame $I$ and its gradients $G$, the depth map $D$, the normal–view inner product map $Nv$, and the warped historical feature $\tilde{h}$ from the previous frame, which is aligned to the current freame. The refinement process yields an updated historical feature representation for the current frame:
\begin{equation}
h = \operatorname{RefineNet}(G, D, Nv, \tilde{h}).
\end{equation}

Finally, the current historical feature $h$ is utilized to generate a residual feature $\Delta F$ via a convolutional operation:
\begin{equation}
\Delta F = \operatorname{Conv}(h),
\end{equation}
and the updated feature is then obtained as $F \leftarrow F + \Delta F$, which follows the formulation in Eq.~\eqref{eq:ctc}.

As mentioned above, $F$ represents the coefficient vector in the polynomial space. The refinement network updates $F$ by leveraging both current-frame features and temporally propagated information, resulting in more robust and temporally consistent outputs. This design provides two main advantages: (1) it remains lightweight, ensuring computational efficiency, and (2) it effectively leverages the geometric and temporal information available in the current frame, thereby enhancing overall interpolation accuracy.

\textbf{Loss Function Design.}
To ensure both accurate super sampling and faithful detail reconstruction, the loss function for GASS consists of two components: the Charbonnier loss~\cite{bruhn2005lucas} $\mathcal{L}_C$ provides the primary photometric constraint, and the Laplace loss $\mathcal{L}_{{lap}}(X, Y) = \left\Vert \nabla^2 X - \nabla^2 Y \right\Vert_1$ emphasizes fine-grained detail consistency. The overall loss function for GASS is formulated as:
\begin{equation}
\mathcal{L} = \mathcal{L}_C + \lambda_1 \mathcal{L}_{{lap}},
\end{equation}
where $\lambda_1 = 0.1$ is weighting factor.


\subsection{Lightweight Temporal Frame Interpolation}

To further enhance rendering efficiency, we introduce a LTFI module that leverages the inherent spatio-temporal coherence of 3DGS. Unlike existing video frame interpolation methods~\cite{kong2022ifrnet, kalluri2023flavr, guo2024generalizable} that employ large and computationally expensive networks, our LTFI is explicitly designed to be \textbf{lightweight and efficient}. It effectively exploits gradient and geometric priors, along with historical temporal information, to achieve real-time interpolation. Given two frames $I_0$ and $I_1$, the objective is to predict the intermediate frame $I_{0.5}$. The network comprises three key components: (1) forward–backward warping, (2) a feature fusion network, and (3) a temporal recurrent unit.

\textbf{Forward–Backward Warping.}
We first warp the preceding high-resolution frame and the subsequent low-resolution frame to the intermediate time step ($t=0.5$). Specifically, the high-resolution image $I_0$, depth map $D_0$, and normal map $N_0$ from frame 0 are warped to the intermediate frame, and the low-resolution image $I_1$, depth map $D_1$, normal map $N_1$, and gradient map $G_1$ from frame 1 are projected in the same manner. The warping operation is guided by motion vectors derived from camera poses. For a pixel $x$ in frame 1, the motion vector is defined as:
\begin{equation}
V_{1 \rightarrow 0.5}[x] = \text{Proj}(\text{UnProj}(x, P_1), P_{0.5}) - x,
\end{equation}
where $P_i$ denotes the camera parameters of frames $i$, respectively. The warped gradient map is computed as:
\begin{equation}
G_{1 \rightarrow 0.5} = \nabla V_{1 \rightarrow 0.5} \cdot w(G_1, V_{1 \rightarrow 0.5}),
\end{equation}
where $w(\cdot)$ denotes the warping operator and $\nabla$ represents spatial differentiation.
We employ high-resolution features from the preceding frame and low-resolution features from the subsequent frame to minimize latency and facilitate parallel processing, thereby preserving temporal consistency and improving computational efficiency.

\textbf{Feature Fusion Network.}
The warped inputs from both directions are encoded by convolutional layers into $F_{0\to0.5}$ and $F_{1\to0.5}$, which are then passed to a Feature Fusion Network (FFN)—a lightweight U-Net–Like architecture illustrated in Fig.~\ref{fig:sr} (b). Each base block integrates max-pooling and upscaling operations to extract low-frequency structures. To maintain efficiency, we incorporate a Spatial Attention Block (SAB)~\cite{wu2024ultralight} into the skip connections to enhance feature discriminability under complex occlusions and suppress irrelevant regions. For upscaling, we employ the pixel shuffle operation to increase spatial resolution via channel rearrangement, which is computationally cheaper and avoids artifacts compared to transposed convolutions.

The FFN outputs a blending weight map $\alpha$ and a fused color prediction $\hat{I}_{0.5}$. The final interpolated high-resolution frame is computed as:
\begin{equation}
I_{0.5} = \alpha \hat{I}_{0.5} + (1 - \alpha) I_{0 \rightarrow 0.5}.
\end{equation}
This adaptive blending formulation balances the reliability of warped content and the accuracy of fused predictions, preserving fine details while mitigating temporal flicker.

\textbf{Temporal Recurrent Unit.}
To further enforce temporal consistency, we introduce a {Temporal Recurrent Unit (TRU)} that explicitly models inter-frame dependencies. The TRU is a modified GRU, which takes as input the warped historical hidden state $h_{0\rightarrow0.5}$ and the current fused feature $F_{0.5}$, producing an updated hidden representation:
\begin{equation}
h_{0.5} = U(h_{0 \rightarrow 0.5}, F_{0.5}),
\end{equation}
where $U(\cdot)$ denotes the recurrent update function. The hidden state $h_{0.5}$ progressively accumulates image features and geometric priors from previous frames, forming a compact spatio-temporal representation of the scene. This design facilitates consistent motion estimation and appearance interpolation across frames, ensuring smooth and stable rendering in 3DGS scenes.

\textbf{Loss Function Design.}
To improve image detail restoration and boost temporal consistency across image sequences, we design the following loss function. A Charbonnier loss \(\mathcal{L}_C\) \cite{bruhn2005lucas} is employed to improve the robustness of reconstructed images against outliers. A perceptual loss \(\mathcal{L}_{\text{VGG}}\) \cite{johnson2016perceptual} is used to enhance the perceptual quality of reconstructed images through high-level feature constraints. To constrain the adaptive blending process, ensuring that it restores only unreliable regions in \(I_{0 \to 0.5}\), we incorporate an \(\mathcal{L}_2\) regularization term on \(\alpha\), defined as \(\mathcal{L}_\alpha = \| \alpha \|_2\). To further mitigate inter-frame flickering, an occlusion-aware loss is additionally introduced, formulated as follows:
\begin{equation}
\mathcal{L}_{\text{OCC}} = \sum_{j=0,1} \mathcal{L}_1(w(I_{0.5}, V_{0.5\to j}), I_{j}).
\end{equation}
Finally, the overall loss function is formulated as:
\begin{equation}
\mathcal{L} =  \mathcal{L}_C + \lambda_1 \mathcal{L}_{\text{VGG}} + \lambda_2 \mathcal{L}_{\text{OCC}} + \lambda_3 \mathcal{L}_{\alpha},
\end{equation}
where \(\lambda_1\!=\!\lambda_3\!=\!0.2\) and \(\lambda_2\!=\!0.05\) are weighting~factors.

\section{Experiment}

\subsection{Implementation Details}

Our method is implemented based on the 3DGS framework~\cite{kerbl20233d} and employs RaDe-GS~\cite{zhang2024rade} for rendering depth and normal maps. Gradient computation, motion vector estimation, and warping operations are implemented through custom CUDA kernels. Our model is trained with the Adam optimizer at a learning rate of $10^{-3}$. For inference, the network is deployed with TensorRT for accelerated execution, where super sampling, and frame interpolation are parallelized to achieve optimal efficiency. All experiments are conducted on an NVIDIA RTX 3090 GPU.

\subsection{Datasets and Metrics}

We evaluate our method on three widely used public benchmarks: Mip-NeRF 360~\cite{barron2022mipnerf360}, Tanks and Temples~\cite{Knapitsch2017}, and Deep Blending~\cite{hedman2018deep}. These datasets cover diverse scenarios, ranging from unconstrained outdoor environments to complex indoor scenes, thereby offering substantial diversity. Our model is trained on the \textit{Garden} scene from Mip-NeRF 360 and evaluated on all available scenes, with average performance reported over 300-frame sequences.

We use standard PSNR and SSIM as quantitative metrics. To assess runtime efficiency, we also report the average FPS. Since our method operates as a post-rendering enhancement, the ground-truth images are generated using standard 3DGS~\cite{kerbl20233d} renderings. This setup ensures that all quality differences stem exclusively from the super sampling or frame interpolation algorithms, rather than discrepancies in the underlying 3D reconstruction, thereby facilitating a fair comparison with other super sampling and frame interpolation methods. To further validate the effectiveness of our approach under real-world conditions, we also include an evaluation using captured real images as ground truth.

To thoroughly evaluate the effectiveness of our approach, we compare against the following categories of methods: (1) \textbf{Traditional Interpolation}: Bicubic interpolation; (2) \textbf{Learning-based Super Sampling}: NinaSR-B1~\cite{ninasr}, RT4KSR~\cite{Zamfir_2023_CVPR}, and ECBSR~\cite{zhang2021edge}; (3) \textbf{Video Frame Interpolation}: FLAVR~\cite{kalluri2023flavr}, InterpAny~\cite{zhong2024clearer}, and GIMM-VFI~\cite{guo2024generalizable}; and (4) \textbf{3DGS Acceleration}: Mini-Splatting~\cite{fang2024mini}, Speedy-Splat~\cite{hanson2025speedy}, and AdR-Gaussian~\cite{xzwang2024adrgaussian}.

\begin{figure*}[t] 
\centering 
\includegraphics[width=1.0\textwidth]{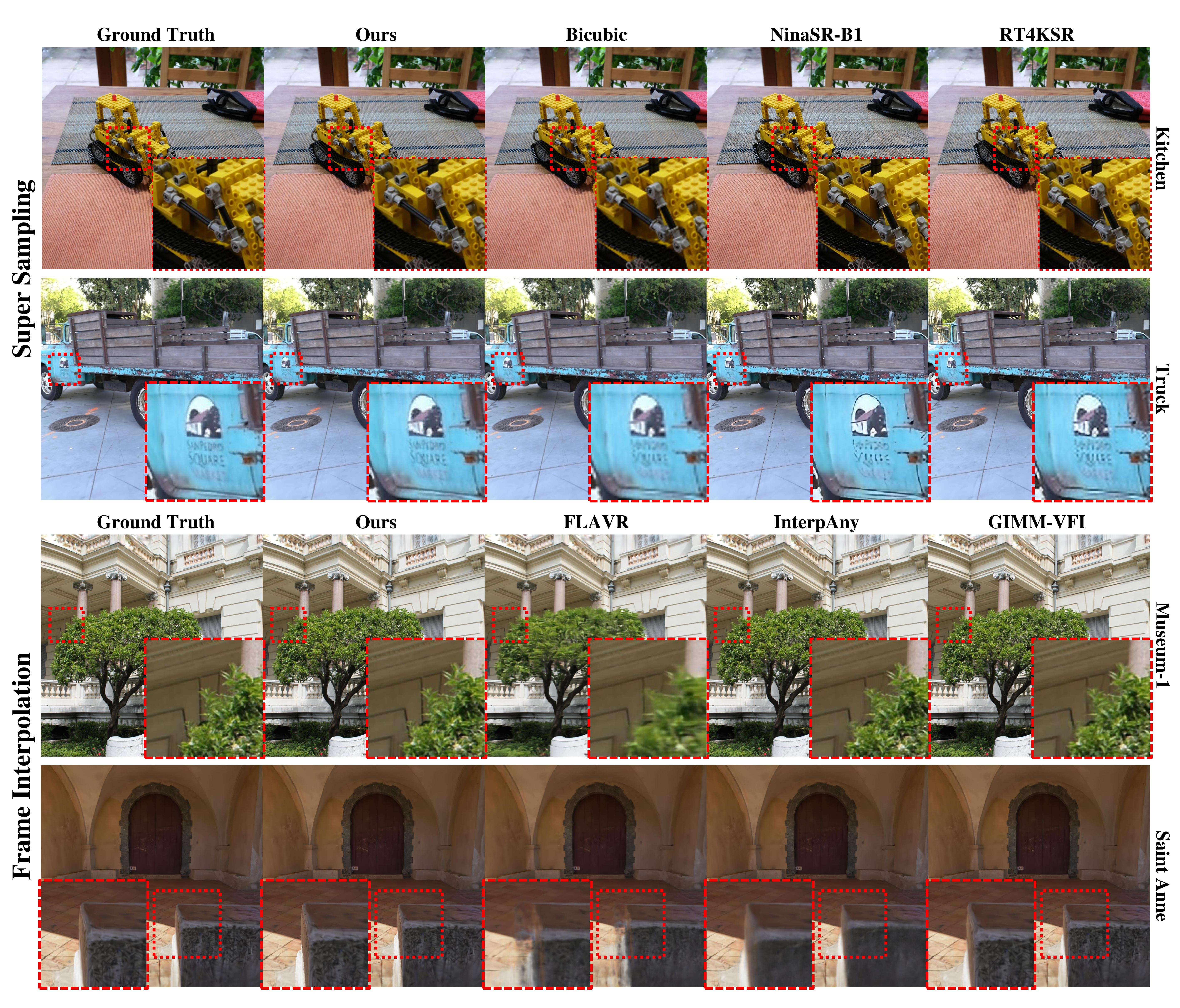}
\caption{Qualitative comparison of super sampling and frame interpolation results. Our method achieves superior detail preservation in super sampling and smoother temporal consistency in frame interpolation compared to existing approaches.}
\label{fig:cmp}
\end{figure*}

\begin{table*}[t]
	\caption{Quantitative comparisons on the Mip-NeRF 360~\cite{barron2022mipnerf360}, Tanks and Temples~\cite{Knapitsch2017}, and Deep Blending~\cite{hedman2018deep} datasets.  All super-resolution methods employ a \(4\times\) scale. Our method achieves the highest FPS among all approaches while maintaining highly competitive visual quality.}
    \centering
	\footnotesize
    \resizebox{\columnwidth}{!}
    {
	\begin{tabular}{l|ccc|ccc|ccc}
		\toprule
		\multirow{3}{*}{Method} & \multicolumn{3}{c|}{MipNeRF 360} & \multicolumn{3}{c|}{Deep Blending} & \multicolumn{3}{c}{Tanks and Temples}  \\
        & \multicolumn{3}{c|}{\(4096\times 2160\)} & \multicolumn{3}{c|}{\(2560\times 1440\)} & \multicolumn{3}{c}{\(2560\times 1440\)}  \\
		& PSNR↑ & SSIM↑ & FPS↑ & PSNR↑ & SSIM↑ & FPS↑ & PSNR↑ & SSIM↑ & FPS↑ \\\midrule
        3DGS (HR) & ~ -        & ~ -        & 17.35    & ~  -        & ~  -       & 41.57    & ~    -       & ~  -         & 50.40     \\
        Ours         & \textbf{49.22}     & \textbf{0.995}     & \textbf{96.47}    & 40.85      & 0.983     & \textbf{226.50}   & 37.64       & 0.978       & \textbf{246.35}    \\\midrule
        Bicubic      & 47.63     & 0.992     & 84.20    & 36.41      & 0.942     & 149.26   & 35.73       & 0.954       & 165.59    \\
        NinaSR-B1    & 40.49     & 0.985     & 4.47     & 30.41      & 0.901     & 16.60    & 29.88       & 0.923       & 16.77     \\
        RT4KSR       & 43.82     & 0.986     & 12.00    & 33.85      & 0.922     & 39.47    & 32.73       & 0.936       & 40.14     \\
        ECBSR        & 44.97     & 0.989     & 47.30    & 35.83      & 0.940     & 66.73    & 34.73       & 0.951       & 63.28     \\\midrule
        FLAVR        & N/A       & N/A       & N/A      & 34.73      & 0.939     & 0.67     & 31.14       & 0.939       & 0.67      \\ 
        IntepAny     & 36.41     & 0.962     & 1.25     & \textbf{48.43}      & \textbf{0.996}     & 4.93     & \textbf{40.11}       & \textbf{0.992}       & 4.93      \\ 
        GIMM-VFI     & 42.57     & 0.989     & 1.59     & 42.34      & 0.987     & 4.83     & 38.47       & 0.985       & 4.76  \\ 
		\bottomrule
	\end{tabular}
    }
    \label{tab:comparison}
\end{table*}

\begin{table}[!t]
	\caption{Quality comparison of 3DGS methods integrated with our approach. For scenes where the real-captured ground truth images have a maximum resolution lower than \(4096\times 2160\), we upsample them to this target resolution via bicubic interpolation followed by cropping to ensure consistent evaluation.}
	\footnotesize
    \centering
    \resizebox{1\linewidth}{!}{
	\begin{tabular}{l|ccc|ccc|ccc|ccc}
		\toprule
		\multirow{3}{*}{Method} & \multicolumn{6}{c|}{\(2560\times 1440\) } & \multicolumn{6}{c}{\(4096\times 2160\) } \\
		& \multicolumn{3}{c|}{w/o Ours} & \multicolumn{3}{c|}{w/ Ours} & \multicolumn{3}{c|}{w/o Ours} & \multicolumn{3}{c}{w/ Ours}\\
		&  PSNR↑ & SSIM↑ & FPS↑ & PSNR↑ & SSIM↑ & FPS↑ & PSNR↑ & SSIM↑ & FPS↑ & PSNR↑ & SSIM↑ & FPS↑\\\midrule
        3DGS                  & 28.95 & 0.859 &  47.11 & 28.92 & 0.858 & \textbf{188.43} & 30.68 & 0.906 & 17.35 & 30.47 & 0.899 & \textbf{96.47}  \\
		Mini-Splatting        & 28.88 & 0.878 & 181.83 & 28.76 & 0.875 & \textbf{463.90} & 29.88 & 0.895 & 56.00 & 29.66 & 0.891 & \textbf{158.78} \\
        Speedy-Splat          & 28.33 & 0.858 & 146.81 & 28.20 & 0.857 & \textbf{261.81} & 28.33 & 0.858 & 79.93 & 28.20 & 0.857 & \textbf{130.68} \\
        AdR-Gaussian          & 29.68 & 0.886 & 127.10 & 29.67 & 0.885 & \textbf{260.60} & 28.38 & 0.872 & 58.27 & 28.07 & 0.867 & \textbf{128.70} \\
		\bottomrule
	\end{tabular}
    }
    \label{tab:synergy}
\end{table}

\subsection{Comparisons}

Table~\ref{tab:comparison} presents a quantitative comparison of our method against various state-of-the-art approaches across three benchmark datasets. As shown, our method achieves the highest FPS among all super sampling and frame interpolation methods, demonstrating its superior runtime efficiency. Moreover, our approach achieves higher image quality than existing learning-based super sampling methods. 

Our method surpasses all interpolation approaches in terms of FPS by a large margin, while achieving highly competitive visual quality. It is worth noting that InterpAny~\cite{zhong2024clearer} and other interpolation methods attain high quality by employing large networks and leveraging high-resolution ground-truth frames as inputs. In contrast, our lightweight approach processes low-resolution inputs and synthesizes the intermediate frame from a super-resolved previous frame and a low-resolution subsequent frame, without access to ground-truth supervision. Despite this challenging setting, our method produces results with fidelity comparable to these heavyweight techniques. As illustrated in Fig.~\ref{fig:cmp}, the qualitative comparisons confirm that our method produces superior details and smoother temporal transitions in both super sampling and frame interpolation tasks underscoring the effectiveness of our design.

To further validate the efficacy of our method, we conduct an additional evaluation using real-captured images as ground truth. As shown in Table~\ref{tab:synergy}, our method incurs only negligible degradation in PSNR and SSIM compared to the original 3DGS renderings, even under $4 \times$ super-resolution and frame interpolation. These results confirm that our method maintains high visual fidelity while delivering significant rendering acceleration, demonstrating its applicability to real-world deployment.

\subsection{Integration with 3DGS Acceleration Methods}
To further validate the generality and scalability of our approach, we integrate it with several representative 3DGS acceleration methods. Table~\ref{tab:synergy} summarizes the rendering efficiency, measured in FPS, after incorporating our post-rendering module into these acceleration frameworks. The experimental results show that our method consistently improves the rendering efficiency of all evaluated techniques, confirming its strong synergy with diverse acceleration strategies. This integration offers a flexible and robust solution for achieving high-frame-rate 3DGS rendering. Notably, the FPS at 4K resolution remains unchanged due to the pipeline-parallel architecture, where the super sampling and frame interpolation networks currently constitute the main performance bottleneck.




\begin{figure}[t]
  \centering
  \begin{minipage}[t]{0.48\textwidth}
    \centering
    \vspace{-\abovecaptionskip}
    \captionof{table}{Ablation study on key components of our approach.}
    \vspace{\abovecaptionskip}
    \label{tab:ablation}
    \resizebox{0.98\linewidth}{!}{
    \begin{tabular}{l|ccc}
      \toprule
       & PSNR↑ & SSIM↑ & FPS↑ \\\midrule
      FULL      & \textbf{40.85} & \textbf{0.983} & 226.50 \\
      w/o GASS & 40.34 & 0.984 & 75.23  \\
      w/o LTFI & 42.13 & 0.985 & 158.66 \\\midrule
      w/o GAI  & 36.81 & 0.960 & 226.50 \\
      w/o GTRR & 39.39 & 0.976 & 226.50 \\\midrule
      w/o GI   & 39.35 & 0.976 & \textbf{233.85} \\
      w/o TRU  & 39.63 & 0.977 & 227.21 \\
      \bottomrule
    \end{tabular}
    }
  \end{minipage}
  \begin{minipage}[t]{0.48\textwidth} 
    \centering
    \caption{Ablation of GASS, GAI, GI, and TRU components.}
    \includegraphics[width=\linewidth, keepaspectratio]{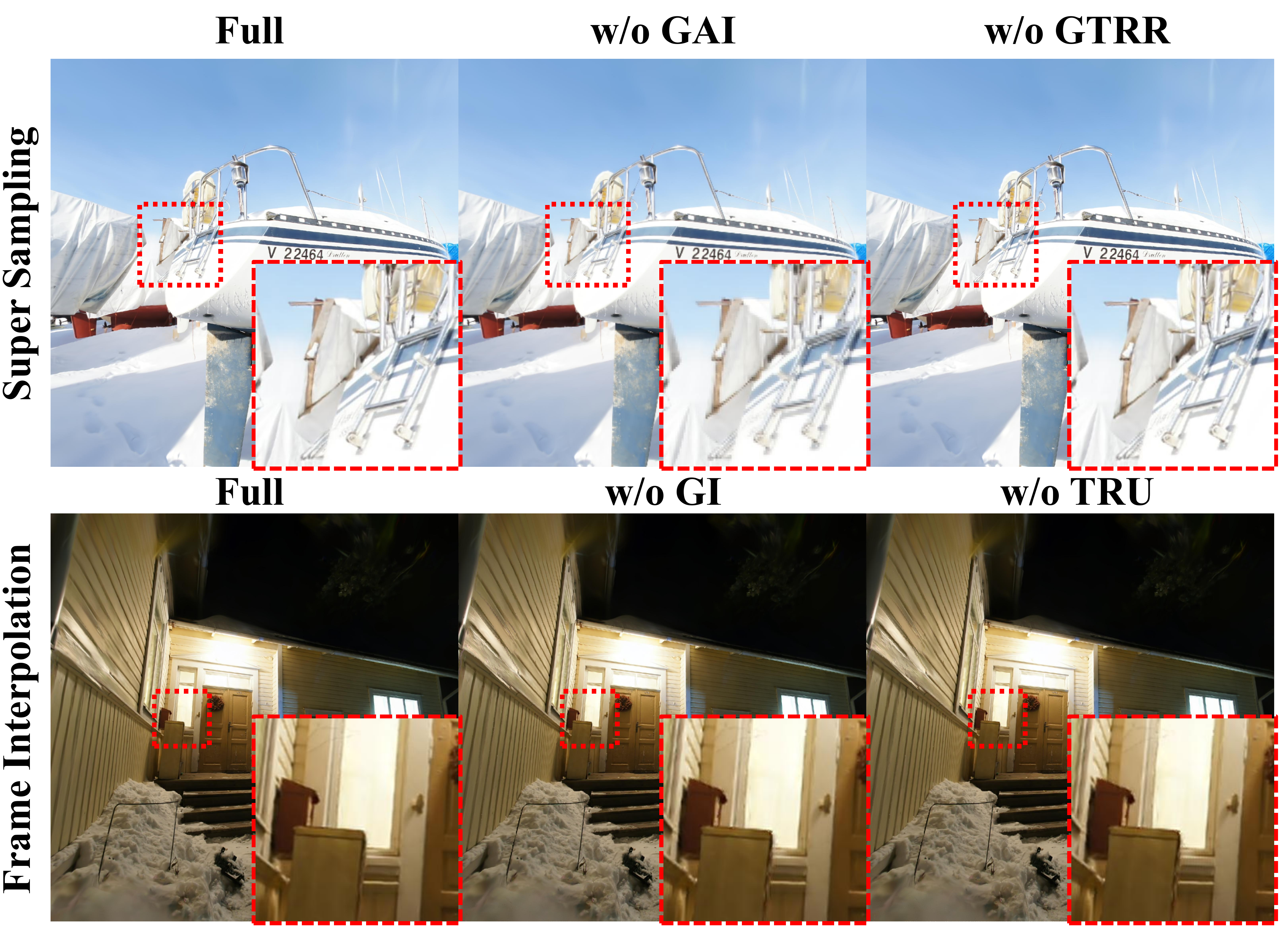}
    \label{fig:ab}
  \end{minipage}

\end{figure}

\subsection{Ablation Studies}
To evaluate the effectiveness of each key component in our approach, we conduct ablation experiments on the Deep Blending~\cite{hedman2018deep} dataset at a resolution of $2560 \times 1440$. We remove the following components and compare them against the full model: (1) without Gradient-Aware Super Sampling (w/o GASS), (2) without Lightweight Temporal Frame Interpolation (w/o LTFI), (3) without Gradient-Aware Interpolation in GASS (w/o GAI), (4) without Geometry-Temporal Recurrent Refinement in GASS (w/o GTRR), (5) without Gradient Information in LTFI (w/o GI), and (6) without Temporal Recurrent Unit in LTFI (w/o TRU).

The quantitative results in Table~\ref{tab:ablation} indicate that neither the super sampling nor the frame interpolation branch alone can achieve optimal performance. For the supe sampling branch, removing GASS or GAI leads to a significant degradation in PSNR and SSIM. For frame interpolation, eliminating GI or TRU slightly improves the frame rate but leads to a noticeable decline in image quality. As illustrated in Fig.~\ref{fig:ab}, the absence of GASS, GAI, GI, or TRU causes blurred textures and visible temporal artifacts, further highlighting the importance of these components.


\section{Conclusion}
In this paper, we present 3DGS$^3$, a novel framework that reexamines 3DGS acceleration from the perspective of post-rendering. Unlike prior methods that focus on optimizing the splatting pipeline itself, our approach introduces a unified framework for super sampling and frame interpolation, enabling the efficient reconstruction of high-resolution and high-frame-rate renderings from low-resolution 3DGS outputs. Specifically, we propose a gradient-aware super sampling module that leverages the spatial gradients of Gaussian splats for high-fidelity detail reconstruction. In addition, a lightweight temporal frame interpolation module based on a compact U-Net-like architecture synthesizes intermediate frames by fusing temporal and differentiable spatial cues from consecutive inputs. Extensive experiments demonstrate that our method outperforms state-of-the-art approaches in both rendering quality and efficiency, while remaining fully compatible with existing 3DGS acceleration techniques, thereby advancing real-time rendering performance with reduced computational overhead. We note that our current method focuses primarily on static scenes; extending it to dynamic 3DGS representations remains a promising direction for future work. Additionally, we aim to extend our framework to mobile platforms by developing compact and model-compressed network variants for efficient on-device deployment.

\clearpage  

%
%
\bibliographystyle{splncs04}
\bibliography{main}
\end{document}